\documentclass[twocolumn,showpacs,preprintnumbers,superscriptaddress]{revtex4}
\usepackage{amsmath}
\usepackage{dcolumn}
\usepackage{bm}
\usepackage{graphicx}

\begin{document}

\title{Quantum information processing architecture with endohedral
fullerenes in a carbon nanotube}
\author{W. L. Yang}
\affiliation{State Key Laboratory of Magnetic Resonance and Atomic and Molecular Physics,
Wuhan Institute of Physics and Mathematics, Chinese Academy of Sciences,
Wuhan 430071, China }
\author{Z. Y. Xu}
\affiliation{State Key Laboratory of Magnetic Resonance and Atomic and Molecular Physics,
Wuhan Institute of Physics and Mathematics, Chinese Academy of Sciences,
Wuhan 430071, China }
\affiliation{Graduate School of the Chinese Academy of Sciences, Beijing 100049, China}
\author{H. Wei}
\affiliation{Center for Modern Physics and Department of Physics, Chongqing University,
Chongqing 400044, China}
\author{M. Feng}
\altaffiliation{mangfeng@wipm.ac.cn}
\affiliation{State Key Laboratory of Magnetic Resonance and Atomic and Molecular Physics,
Wuhan Institute of Physics and Mathematics, Chinese Academy of Sciences,
Wuhan 430071, China }
\author{D. Suter}
\altaffiliation{dieter.suter@uni-dortmund.de}
\affiliation{Fakult$\ddot{a}$t Physik, Technische Universit$\ddot{a}$t Dortmund, 44221
Dortmund, Germany}

\begin{abstract}
A potential quantum information processor is proposed using a fullerene
peapod, i.e., an array of the endohedral fullerenes $^{15}N@C_{60}$ or $%
^{31}P@C_{60}$ contained in a single walled carbon nanotube (SWCNT). The
qubits are encoded in the nuclear spins of the doped atoms, while the
electronic spins are used for initialization and readout, as well as for
two-qubit operations.
\end{abstract}

\maketitle

\section{INTRODUCTION}

Quantum information processing (QIP) has attracted considerable attention
over the last years, since it allows one to efficiently solve computational
problems that have no efficient solution in classical computer science \cite%
{Shor,Grover}. Unleashing this potential requires the design of quantum
information processors that control a large number of qubits. Among the
interesting qubit candidates are endohedral fullerenes $^{15}N@C_{60}$ or $%
^{31}P@C_{60}$ \cite{Sta,Har,Suter,Twa,Feng1,Ju} whose decoherence times are
longer than those of most other candidates. Quantum gate operations on these
qubits can be implemented by electron spin resonance (ESR) and nuclear
magnetic resonance (NMR) techniques.

The endohedral atoms ($^{15}$N or $^{31}$P) have electron spins as well as
nuclear spin degrees of freedom. The electron spins are particularly well
suited for individual addressing, for coupling neighboring qubits, and for
readout. The nuclear spins, in contrast, are not coupled to each other, and
their decoherence time is particularly long. It appears therefore useful to
combine these properties and use the nuclear as well as the electronic spin
for quantum information processing \cite{Har,Suter,Twa,Feng1,Ju}. This
combination can also solve some problems associated with the fact that the
electronic spin is not a two-level system, but has $S=3/2$.

\begin{figure}[tbph]
%\centering
%\includegraphics[width=2.0in]{fig1}
\includegraphics[width=3.2 in]{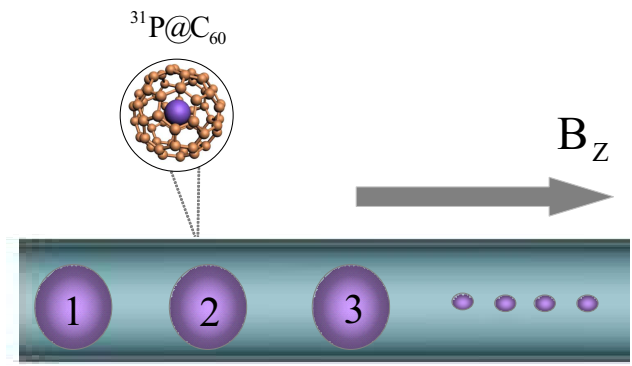}
\caption{(Color online) Schematic setup of $^{31}$P@C$_{60}$@SWCNT system,
where the fullerene molecules $^{31}$P@C$_{60}$ are trapped equidistantly. A
magnetic field gradient, generated by micropatterned wires, is applied. }
\label{fig1}
\end{figure}

One of the main obstacles for implementing such fullerene-based quantum
computers is the lack, until now, of a single-qubit readout capability \cite%
{Feng2, readout}. Here, we would like to discuss a system that might solve
this difficulty. We consider a row of endohedral fullerenes $^{15}$N@C$_{60}$
or $^{31}$P@C$_{60}$ enclosed in a single walled carbon nanotube (SWCNT).
This system has been called fullerene peapod. Such systems have been studied
experimentally \cite{Smi,Pea2}. One of the attractive features of the SWCNT
is that it allows transport of mobile electrons \cite{Pea2}, which could be
used to read out the state of the electron spins in the peapod. This idea is
sketched in Fig. \ref{fig1} where the doped fullerenes are positioned as a
line in a SWCNT. Using the mobile electrons on the SWCNT may help to solve
the problem of the single-spin readout, which was left open in earlier
proposals \cite{Har,Suter,Twa,Feng1,Ju}.

Our paper is structured as follows. In Sec. II we present our system
architecture and work through the requirements for implementing universal
quantum computation \cite{Div}. This includes one-qubit and two-qubit
gating, initialization and readout. Sec. III discusses some details that
arise when large quantum registers are implemented. Section IV gives the
requirements that must be met by such an apparatus and discusses some issues
related to its realization. The paper ends with a short summary in Sec. V.

\section{THE \textbf{MODEL}}

\subsection{System and Hamiltonian}

The system that we consider consists of a linear chain of endohedral
fullerenes confined in a SWCNT, as sketched in Fig.\,\ref{fig1}. The $^{31}$%
P nuclear spins represent the qubits, while the electronic spins provide
ancilla qubits for coupling the nuclear spin qubits to each other. Between
gate operations, the electron spin state does not contain quantum
information and we are free to choose the state most suitable to our
purpose. In the following, we will assume that it is in the $m_s = +3/2$
state.

We label the states of the auxiliary qubits $\left\vert 3/2\right\rangle
_{S}=\left\vert \uparrow \right\rangle _{S}$ and $\left\vert
-3/2\right\rangle _{S}=\left\vert \downarrow \right\rangle _{S}$. The states
of the nuclear spins encoding the qubits are $\left\vert 1/2\right\rangle
_{I}=\left\vert \uparrow \right\rangle _{I}$ and $\left\vert
-1/2\right\rangle_{I}=\left\vert \downarrow \right\rangle _{I}$. The system
is placed in a magnetic field oriented along the peapod axis. The $i^\mathrm{%
th}$ qubit experiences a field
\begin{equation*}
\vec{B}_{i}=[B_{0}+\frac{dB}{dz} z_{i}]\vec{e}_{z},
\end{equation*}
where $dB/dz$ is the magnetic field gradient, $\vec{e}_{z}$ is a unit vector
along the axis of the peapod, $z_i$ is the position of the $i^{th}$ qubit
from the origin, and $B_0$ is the strength of the magnetic field at the
origin.

The electronic spins of neighboring molecules interact by dipole-dipole
couplings. In this scheme, we have neglected the C60-SWCNT interaction \cite%
{Oka,Dub}, which are far from the resonance frequency of the electron spin
under our consideration. Experimentally, the C60-SWCNT interaction is not
sufficiently large to account for the observed scanning tunneling
spectroscopy (STS) images \cite{Dub}. As in relevant papers published
previously, we only take the nearest-neighbor interactions into account and
find for the system Hamiltonian
\begin{eqnarray}
H &=&\sum_{i=1}^{N}\{\Omega _{S}^{i}S_{z}^{i}-\Omega
_{I}^{i}I_{z}^{i}+AS_{z}^{i}I_{z}^{i}\}  \notag \\
&&+\sum_{i=1}^{N-1}D_{i,i+1}S_{z}^{i}S_{z}^{i+1},  \label{e.Ham}
\end{eqnarray}%
where the first two terms are the electronic and nuclear Zeeman splitting,
the third term denotes the hyperfine interaction and the last term
represents the secular part of the magnetic dipolar interaction between the
nearest-neighbor electronic spins. The electron spin Larmor frequency is $%
\Omega _{S}^{i}=g\mu _{B}B_{i}$, the nuclear spin Larmor frequency $\Omega
_{I}^{i}=\gamma _{I}B_{i}$, $\mu _{B}$ is the Bohr magneton, $g$ is the Land%
\'{e} g-factor of the electron, $\gamma _{I}$ is the gyromagnetic ratio of
the nucleus, $A$ the hyperfine coupling strength, and $D_{i,i+1}$ is the
nearest-neighbor dipole-dipole coupling strength. We have omitted the
couplings between nuclear spins because they are too small and use frequency
units, where $\hbar =1$. We have truncated the dipolar coupling operator
with respect to the difference of the electron spin Larmor frequencies. The
validity of this approximation depends on the magnetic field gradient and on
the distance between the qubits. For a nearest-neighbor distance of $r=2.91$
nm and a magnetic field gradient of $4\times 10^{5}$ T/m, the relevant
resonance frequencies of the electron spin $\left\vert 3/2\right\rangle
\leftrightarrow \left\vert -3/2\right\rangle $ transitions differ by $97.2$ $%
MHz$, while the coupling strength is of the order of 3 $MHz$. Under these
conditions, the truncation is therefore an excellent approximation.

\subsection{Quantum gate operations for a single fullerene}

\label{s.singleF}

\begin{figure}[tbph]
\centering\includegraphics[width=0.7\columnwidth]{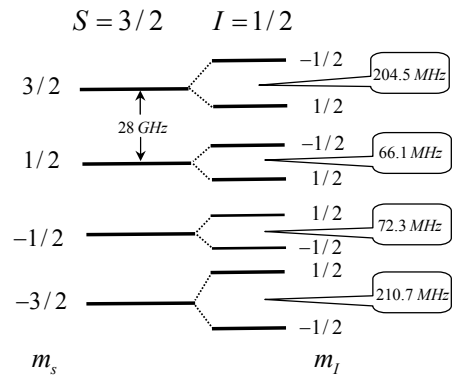}
\caption{Energies of a single endohedral fullerene $^{31}$P@C$_{60}$. }
\label{fig2}
\end{figure}

Consider a single C$_{60}$ molecule in a SWCNT under an external magnetic
field $B_{0}$. The Hamiltonian of Eq. (\ref{e.Ham}) reduces then to
\begin{equation}
H_{1}=\Omega _{S}^{0}S_{z}-\Omega _{I}^{0}I_{z}+AS_{z}I_{z},  \label{e.Ham1}
\end{equation}%
where $\Omega _{S}^{0}=g\mu _{B}B_{0}$ and $\Omega _{I}^{0}=\gamma _{I}B_{0}$%
. For $^{31}$P and $^{15}$N, the hyperfine interaction strength $A/2\pi $
are about $138.4$ $MHz$ and 21.2 $MHz$, respectively \cite{Sta}. For most of
the specific data in the paper, we will use the number from $^{31}$P, but
the concept is equally applicable to both molecules. In a field of $B_{0}=1$
T, the Larmor frequencies of the electron and the $^{31}$P nuclear spins are
$\Omega _{S}^{0}/2\pi \approx 28$ $GHz$ and $\Omega _{I}^{0}/2\pi \approx
17.2$ $MHz$, respectively. Straightforward calculations yield the energies
of the eigenstates as listed in Table I.

\begin{table}[tbp]
\caption{Energies of the spin Hamiltonian (\protect\ref{e.Ham1}). }%
\begin{tabular}{lll}
\hline\hline
\ Nuclear spin & $\ \ $Electron spin & \multicolumn{1}{l}{$\ \ \ \ $Energy \
} \\ \hline
$\ \ \ \ \ \left\vert 1/2\right\rangle _{I}$ & $\ \ \ \ \ \left\vert
3/2\right\rangle _{S}$ & \multicolumn{1}{l}{$\
3\Omega^0_S/2+\Omega^0_I/2-3A/4$} \\
$\ \ \ \ \ \left\vert 1/2\right\rangle _{I}$ & $\ \ \ \ \ \left\vert
1/2\right\rangle _{S}$ & \multicolumn{1}{l}{$\ \Omega^0_S/2+\Omega^0_I/2-A/4$%
} \\
$\ \ \ \ \ \left\vert 1/2\right\rangle _{I}$ & $\ \ \ \ \ \left\vert
-1/2\right\rangle _{S}$ & \multicolumn{1}{l}{$-\Omega^0_S/2+\Omega^0_I/2+A/4$%
} \\
$\ \ \ \ \ \left\vert 1/2\right\rangle _{I}$ & $\ \ \ \ \ \left\vert
-3/2\right\rangle _{S}$ & \multicolumn{1}{l}{$-3\Omega^0_S/2+%
\Omega^0_I/2+3A/4$} \\
$\ \ \ \ \ \left\vert -1/2\right\rangle _{I}$ & $\ \ \ \ \ \left\vert
3/2\right\rangle _{S}$ & \multicolumn{1}{l}{$\
3\Omega^0_S/2-\Omega^0_I/2+3A/4$} \\
$\ \ \ \ \ \left\vert -1/2\right\rangle _{I}$ & $\ \ \ \ \ \left\vert
1/2\right\rangle _{S}$ & \multicolumn{1}{l}{$\ \Omega^0_S/2-\Omega^0_I/2+A/4$%
} \\
$\ \ \ \ \ \left\vert -1/2\right\rangle _{I}$ & $\ \ \ \ \ \left\vert
-1/2\right\rangle _{S}$ & \multicolumn{1}{l}{$-\Omega^0_S/2-\Omega^0_I/2-A/4$%
} \\
$\ \ \ \ \ \left\vert -1/2\right\rangle _{I}$ & $\ \ \ \ \ \left\vert
-3/2\right\rangle _{S}$ & \multicolumn{1}{l}{$-3\Omega^0_S/2-%
\Omega^0_I/2-3A/4$} \\ \hline\hline
\end{tabular}%
\end{table}

\begin{figure}[tbph]
\centering\includegraphics[width=3.2in]{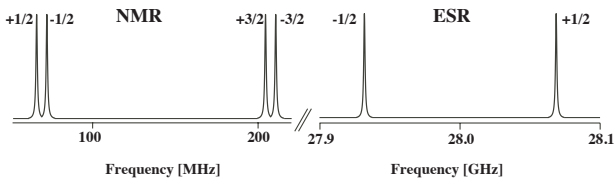}
\caption{Magnetic-dipole transitions for a single endohedral fullerene $%
^{31} $P@C$_{60}$. The left-hand spectrum corresponds to the nuclear spin
transitions (NMR), the right-hand part to the electron spin transitions
(ESR). Each transition of the nuclear spin is labelled with the
corresponding $m_{S}$ value of the electron spin and the electron spin
transitions are labelled with the nuclear spin states. }
\label{spectra}
\end{figure}

Fig. \ref{fig2} represents the energies of the spin states and identifies
them by the spin states. Fig. \ref{spectra} shows the magnetic dipole
transitions between these states, which correspond to a change of the
nuclear spin quantum number by one unit (left hand side, NMR) or the
electron spin (right hand side, ESR). The electron spin transitions have the
frequencies $\Omega _{S}^{0}\pm A/2$ and fall therefore into the microwave
frequency range ($\approx 28$ $GHz$ in a field of 1 T). The nuclear spin
transition frequencies are $3A/2\pm \Omega _{I}^{0}$ and $A/2\pm \Omega
_{I}^{0}$ and fall into the radiofrequency (RF) range ($\approx 70-210$ $MHz$%
).

Resonant microwave pulses are well suited for generating quantum gate
operations in spin qubits. However, if we apply them to the $S=3/2$ electron
spin, they usually mix the (ancilla-) qubit states $\left\vert \pm
3/2\right\rangle _{S}$ with the unused $\left\vert \pm 1/2\right\rangle _{S}$
states, thus causing loss of quantum information. This is not the case for $%
\pi $-rotations around an axis in the $xy$-plane, so we only use this type
of operations on the electron spins. In fact, we have used the decoulping
methods described in Ref. \cite{Suter,Twa}. It is easy to verify that the
flip of the electronic spin at some times can also be implemented on the
electron spin $\frac{3}{2}$ by the inversion operation

\begin{equation*}
\hat{P}=e^{-i\pi S_{x}}=i%
\begin{pmatrix}
0 & 0 & 0 & 1 \\
0 & 0 & 1 & 0 \\
0 & 1 & 0 & 0 \\
1 & 0 & 0 & 0%
\end{pmatrix}%
\end{equation*}%
to achieve%
\begin{equation*}
e^{-i\pi S_{x}}S_{z}e^{i\pi S_{x}}=-S_{z},
\end{equation*}%
where $\hat{P}$ works independently in the subspace spanned by $\left\vert
\pm 3/2\right\rangle _{E}$ or by $\left\vert \pm 1/2\right\rangle _{E}$, we
may employ it to accomplish logic gating between electronic and nuclear
qubits.

In the following, we will need to SWAP information between the electron and
nuclear spin. In terms of spin states, this corresponds to
\begin{equation*}
\left\vert 3/2\right\rangle _{S}\left\vert -1/2\right\rangle
_{I}\leftrightarrow \left\vert -3/2\right\rangle _{S}\left\vert
1/2\right\rangle _{I},
\end{equation*}%
which cannot be induced by a single RF or microwave pulse. It can, however,
be decomposed into CNOT operations:
\begin{eqnarray}
\mathrm{SWAP}_{SI} &=&\mathrm{CNOT}_{SI}\cdot \mathrm{CNOT}_{IS}\cdot
\mathrm{CNOT}_{SI},  \notag \\
&=&\mathrm{CNOT}_{IS}\cdot \mathrm{CNOT}_{SI}\cdot \mathrm{CNOT}_{IS},
\label{e.SWAP}
\end{eqnarray}%
where CNOT$_{SI}$ (CNOT$_{IS}$) is the controlled-NOT operation acting on
the nuclear (electron) spin, controlled by the electron (nuclear) spin. The
first decomposition corresponds to the exchange of the states
\begin{eqnarray}
\left\vert 3/2\right\rangle _{S}\left\vert -1/2\right\rangle _{I}
&\leftrightarrow &\left\vert 3/2\right\rangle _{S}\left\vert
1/2\right\rangle _{I}  \notag \\
\left\vert -3/2\right\rangle _{S}\left\vert 1/2\right\rangle _{I}
&\leftrightarrow &\left\vert 3/2\right\rangle _{S}\left\vert
1/2\right\rangle _{I}  \notag \\
\left\vert 3/2\right\rangle _{S}\left\vert -1/2\right\rangle _{I}
&\leftrightarrow &\left\vert 3/2\right\rangle _{S}\left\vert
1/2\right\rangle _{I}.  \label{e.SWAPSI}
\end{eqnarray}%
Each of these CNOT gates can be implemented by a selective microwave or RF
pulse: The CNOT$_{SI}$, e.g., corresponds to a $\pi $ rotation of the
nuclear spin, conditional on the electron spin being in the $m_{S}=+3/2$
state. As can be seen from Figs. \ref{fig2} and \ref{spectra}, this can be
implemented by a RF pulse with a frequency of 204.5 $MHz$ (in a field of 1
T).

We are now left with the task of implementing arbitrary single-qubit
operations on the nuclear spin qubit. This can in principle be accomplished
with the help of resonant RF pulses, as in NMR quantum computing \cite%
{Va,Su2}. However, these pulses would have to compete with the hyperfine
coupling. Direct application of an RF field that is strong enough to make
the effect of the hyperfine coupling negligible would not only be
technically very challenging, it would also strongly violate the rotating
wave approximation and thereby lead to a very poor fidelity of the gate
operation. Instead, it will be much easier and result in higher fidelity, if
the single-qubit operations are implemented by pairs of RF pulses applied
simultaneously at the frequencies 204.5 $MHz$ and 210.7 $MHz$. One of these
RF fields will implement the operation conditional on the electron spin
being in the logical 0 state, the other if the electron spin is in the
logical 1 state. The combined effect thus corresponds to an unconditional
gate operation.

\subsection{Extended quantum registers}

\label{s.SWAP2} 
\begin{figure}[tbph]
\centering\includegraphics[width=0.7\columnwidth]{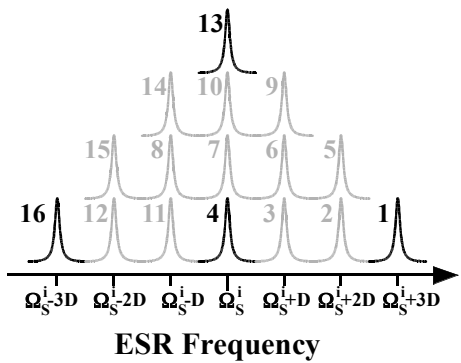}
\caption{Schematic representation of the transition frequencies of the
electron spins in an linear chain of fullerene qubits. The central
transition frequency $\Omega _{S}^{i}$ corresponds to the frequency of an
isolated electron spin and includes contributions from the magnetic field
gradient and the nuclear hyperfine interaction. $D$ is the coupling
constant. Transitions between states that are not populated in an ideal
quantum processor are gray. }
\label{f.fre}
\end{figure}

Large and therefore powerful quantum registers can be
implemented by extended chains of qubits. The different qubits can be
addressed in frequency space by applying a magnetic field gradient. For the
Hamiltonian of Eq. (\ref{e.Ham}), the transition frequencies of every
electron spin split into 16 transitions, whose frequencies depend on the
state of the neighboring electron spin, and which are up to fourfold
degenerate, as shown in Figure \ref{f.fre} and Table \ref{t.Freq}.

\begin{table}[tbp]
\caption{List of the transition frequencies of the electronic spin vs. the
corresponding states of the neighboring electron spins. Transitions that
occur in an ideal quantum information processor are in boldface.}
\label{t.Freq}%
\begin{tabular}{ll|l|l|}
\hline
& \# & state of neighbor qubits & transition frequency \\ \hline
& \textbf{1} & $\mathbf{\quad \left\vert 3/2,3/2\right\rangle}$ & \quad $%
\mathbf{\Omega_S^i+3D}$ \\
& 2 & $\quad \left\vert 3/2,1/2\right\rangle$ & \quad $\Omega_S^i+2D$ \\
& 3 & $\quad \left\vert 3/2,-1/2\right\rangle$ & \quad $\Omega_S^i+D$ \\
& \textbf{4} & $\mathbf{\quad \left\vert 3/2,-3/2\right\rangle}$ & \quad $%
\mathbf{\Omega_S^i}$ \\
& 5 & $\quad \left\vert 1/2,3/2\right\rangle$ & \quad $\Omega_S^i+2D$ \\
& 6 & $\quad \left\vert 1/2,1/2\right\rangle$ & \quad $\Omega_S^i+D$ \\
& 7 & $\quad \left\vert 1/2,-1/2\right\rangle$ & \quad $\Omega_S^i$ \\
& 8 & $\quad \left\vert 1/2,-3/2\right\rangle$ & \quad $\Omega_S^i-D$ \\
& 9 & $\quad \left\vert -1/2,3/2\right\rangle$ & \quad $\Omega_S^i+D$ \\
& 10 & $\quad \left\vert -1/2,1/2\right\rangle$ & \quad $\Omega_S^i$ \\
& 11 & $\quad \left\vert -1/2,-1/2\right\rangle$ & \quad $\Omega_S^i-D$ \\
& 12 & $\quad \left\vert -1/2,-3/2\right\rangle$ & \quad $\Omega_S^i-2D$ \\
& \textbf{13} & $\mathbf{\quad \left\vert -3/2,3/2\right\rangle}$ & \quad $%
\mathbf{\ \Omega_S^i}$ \\
& 14 & $\quad \left\vert -3/2,1/2\right\rangle$ & \quad $\Omega_S^i-D$ \\
& 15 & $\quad \left\vert -3/2,-1/2\right\rangle$ & \quad $\Omega_S^i-2D$ \\
& \textbf{16} & $\mathbf{\quad \left\vert -3/2,-3/2\right\rangle}$ & \quad $%
\mathbf{\ \Omega_S^i-3D}$ \\ \hline
\end{tabular}%
\end{table}

As discussed in section \ref{s.singleF}, single-qubit operations can be
implemented directly on the nuclear spin. Since the nuclear spin transitions
are the same for qubits in extended chains as for isolated qubits, all the
nuclear spin operations remain the same. We do, however, have to modify the
electron spin operations, e.g. the selective inversion of the electron spin
discussed in the second line of Eq. (\ref{e.SWAPSI}). The single resonance
line that has to be inverted in the case of an isolated qubit corresponds
now to sixteen lines at seven frequencies. Single-qubit gate operations
applied to qubit $i$ have to excite all these transitions equally, since
they must not depend on the states of the neighboring qubits. This is a
rather challenging task, which can be implemented by multi-frequency
excitation \cite{Borbat}. The task can be simplified if we assume that our
quantum information processor works under ideal conditions: in this case,
only the $\pm 3/2$ states are occupied, and we can disregard all transitions
in which one of the neighboring qubits is in the states $\pm 1/2$. These
transitions are grayed in Fig. \ref{f.fre} and in normal typeface in Table %
\ref{t.Freq}. This leaves us with 4 transitions at frequencies $\Omega_S^i,
\Omega_S^i\pm3D$, as shown in bold in Table \ref{t.Freq} and as black lines
in Fig. \ref{f.fre}.

\subsection{Two-qubit gates}

We now consider two-qubit operations between neighboring qubits ($i-1$, $i$)
in an extended chain of fullerene molecules. All such operations start with
SWAP$_{SI}$ operations on the molecules participating in the two-qubit
operation. The central part of our two-qubit gate consists of the CNOT$%
_{i-1~i}$ gate or the controlled phase flip CPF$_{i-1~i}$ acting on a
nearest-neighbor pair of electron spins. The gate ends with a second pair of
SWAP$_{SI}$ operations on the molecules participating in the two-qubit
operation. The net effect of this sequence of operations is that the two
nuclear spins have undergone the CNOT$_{i-1~i}$ (or CPF$_{i-1~i}$)
operation, while the electron spins return to their initial state (which we
assume to be the $m=+3/2$ ground state). This assures that the passive
electron spins are always in the $m=+3/2$ state, as we assumed in the
preceding subsection. Furthermore, it minimizes the effect of decoherence,
since this is the thermal equilibrium state at low temperature and therefore
not affected by relaxation.

Considering now the central part of the two-qubit operation, the CNOT$%
_{i-1,i}$ gate between the electron spins, inspection of Table \ref{t.Freq}
shows that it can be implemented by applying $\pi $ pulses to transitions 1
and 4: This implements a operation NOT on qubit $i$ conditional on the qubit
$i-1$ being in state +3/2 and unconditional with respect to the state of
qubit $i+1$. Unfortunately, a $\pi $-pulse applied to transition 4 will also
invert transition 13, which is degenerate with transition 4, and which
corresponds to qubit $i-1$ being in state $-3/2$ and therefore invalidates
the CNOT operation. This difficulty can be eliminated by various approaches,
including multifrequency excitation of several transitions. Alternatively,
we may take into account that qubit $i+1$ is not active during a CNOT$%
_{i-1,i}$ operation, i.e. the information is in the nuclear spin, and the
state of the electron spin is the state into which it was initialized. Here,
we assume that this is the +3/2 state. This eliminates the need to invert
transition 4 and makes the $\pi $ pulse on transition 1 a perfect CNOT$%
_{i-1,i}$ operation.

In an extended quantum register, we also have to consider that the
operations on the nuclear spins will affect all nuclei in the quantum
register, i.e. also passive qubits, which should not be affected by the
specific gate operations. If we try to make the pulses on the nuclear spins
sufficiently weak, so that their effect on the passive qubits is negligible,
they will become unacceptably slow: with the parameters as discussed above,
the difference in the Larmor frequencies of neighboring nuclear spins is
only on the order of 10 $Hz$ (for $^{31}$P). It will then be advantageous to
use hard pulses that simultaneously affect all nuclear spins. Since the only
operations applied to the nuclear spins in our scheme are the SWAP$_{SI}$
operations of Eq. (\ref{e.SWAP}), this turns out not to be a problem: For
the passive qubits, where the CNOT$_{IS}$ operations are identity
operations, the first version of the SWAP$_{SI}$ gate becomes CNOT$_{SI}$
CNOT$_{SI}$ = 1, while the second version becomes CNOT$_{SI}$. Since all
operations (one- and two-qubit) involve pairs of SWAP$_{SI}$ gates, even the
second version leaves passive qubits invariant.

\subsection{Initialization and readout}

\label{s.readout}

The proposed system can be initialized into the ground state by relaxation
of the electron spins: In a field of $B=1$T, at a temperature $T=0.1$K, the
equilibrium ground state population is $>0.99999$. This electron spin
polarization can be transferred to the nuclear spin qubits by a SWAP
operation.

Readout may be accomplished by using mobile electrons on the outside of the
peapod. Various schemes have been developed for the detection of single
spins \cite{Det,Ma,Pake,Park}, which cannot be applied directly to the spins
of electrons in C$_{60}$ cages. Nevertheless, this can also be achieved, by
coupling the caged spin to external spins and converting the spin degree of
freedom to charges, then detecting the charges \cite{Feng2,readout}.

\begin{figure}[tbph]
\centering\includegraphics[width=3.5 in]{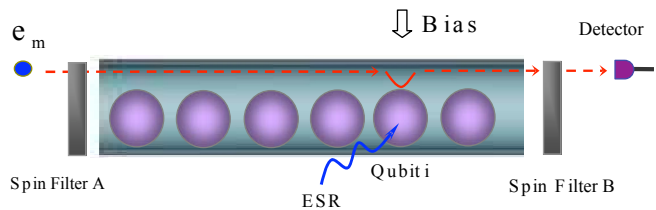}
\caption{Scheme for the detection of the state of electron-spins inside a
fullerene embedded in a SWCNT. The spin filter $A(B)$ located in the front
(back) of the SWCNT only allows the down (up)-polarized electrons to pass.
The bias stops the mobile electron along the SWCNT at a place above the
caged spin and an ESR pulse generates a CNOT operation between the mobile
and the caged electron spins.}
\label{fig4}
\end{figure}

This can also be achieved in the present system, as shown schematically in
Fig. \ref{fig4}. The electrons used for detection are spin-polarized by
passing them through a spin filter and guiding them along the SWCNT by
applying suitable bias fields. At the position of the qubit that is to be
read out, the mobile electron is stopped by adjusting the bias field. The
separation between the mobile and the caged electrons is small enough that a
sizable magnetic dipole-dipole coupling exists between them. We may then
write the Hamiltonian of the two electron spins as
\begin{equation}
H_{sub}=\Omega _{S}^{i}S_{z}^{i}+\Omega _{S}^{m}S_{z}^{m}+D^{^{\prime
}}S_{z}^{i}S_{z}^{m},  \label{e.mobcoupl}
\end{equation}%
where $S_{z}^{i}$ is the spin operator of the caged qubit electron ($S=3/2$)
and $S_{z}^{m}$ the spin of the mobile electron ($S=1/2$) and $D^{^{\prime
}} $ represents the strength of the coupling between the mobile and the
caged spins. For a separation of 0.8 nm, the dipole coupling strength would
be of the order of 100 $MHz$. Depending on the state of the caged electron ($%
m=\pm 3/2$), the transition frequency of the mobile electron is thus shifted
from its Larmor frequency $\Omega _{S}^{m}$ by $\pm 3/2D^{^{\prime }}$.

This can be used for a detection mechanism: If we want to detect the $m=+3/2$
state of the caged electron, we send a stream of electrons along the SWCNT,
which are spin-polarized in the spin filter A. This spin filter can be made
from ferromagnetic materials \cite{Mo}, or semiconductor quantum dots \cite%
{Eng}. When the electrons are trapped near qubit $i$, we apply a microwave
pulse with frequency $\Omega _{S}^{m} + 3/2 D^{^{\prime}}$ and a flip angle
of $\approx \pi$. This operation corresponds to a CNOT operation conditional
on the caged spin being in the $m=+3/2$ state and thus to a COPY operation.
The mobile spin is then transported further down along the SWCNT and through
spin filter B, which is oriented such that it only allows those electrons to
pass whose spin has been flipped. This procedure can be repeated as often as
necessary to get a sufficient signal at the detector.

\section{Beyond the nearest-neighbor approximation}

\subsection{Non-nearest neighbor couplings}

So far, we have considered only interactions between nearest neighbors. In
reality, dipolar interactions exist between any pair of spins. In a linear
chain, the strength $D_{i,k}$ of the interaction between qubits $i$ and $k$
decreases as
\begin{equation}
D_{i,k}\propto \left( \frac{1}{|i-k|}\right) ^{3}
\end{equation}

\begin{figure}[tbph]
\centering\includegraphics[width=\columnwidth]{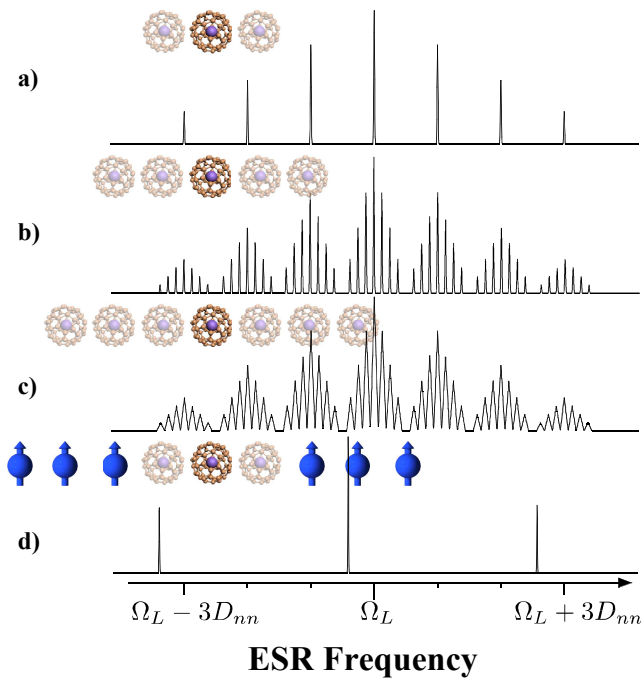}
\caption{Transitions of a single qubit in the presence of couplings to other
qubits as a function of the chain length. a) Nearest neighbors only. b)
Nearest neighbors and next-nearest neighbors. c) Three qubits on each side.
d) Infinite chain, where all but the nearest neighbors are polarized. $%
\Omega _{L}$ is the Larmor frequency of the central qubit and $D_{nn}$ is
the nearest-neighbor coupling strength.}
\label{f.chainspec}
\end{figure}

Figure \ref{f.chainspec} illustrates the effect of the non-nearest neighbor
couplings. The uppermost spectrum corresponds to the case of nearest
neighbors only, i.e. to the situation discussed in section II. If we add
non-nearest neighbors (trace b)), each resonance line splits into 7 lines. A
third pair of qubits (trace c)) results in a line broadening, since the
additional splittings are not resolved in the figure. This situation can be
corrected if we take into account the fact that all electron spins of the
passive qubits are in the $\uparrow (m_S =+3/2)$ state, since the quantum
information was SWAPed into the nuclear spin. This situation is depicted in
trace d), where all but the nearest neighbor qubits are taken to be in the $%
\uparrow$ state. As a result, we do not obtain a splitting of the resonance
line, but a shift by $-3D_{nn}/7 \approx -0.429 D_{nn}$, where $D_{nn}$ is
the coupling between nearest neighbors.

\subsection{Quantum state transfer}

Universal quantum computation requires two-qubit operations between all
qubit pairs. If the pair is not directly coupled by dipolar interaction,
this usually requires a series of SWAP operations between nearest neighbors.
Our present schemes allows us to avoid this overhead by using a mobile
electron. As for the readout, we assume that this mobile electron travels in
the conduction band of the semiconducting carbon nanotube hosting the
fullerenes. Its motion is controlled by external bias electrodes.

\begin{figure}[tbph]
\centering\includegraphics[width=7 cm]{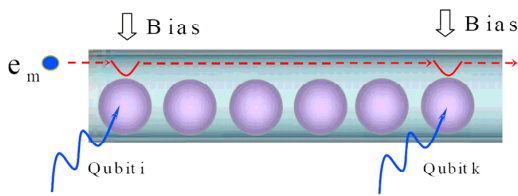}
\caption{Scheme for transferring a quantum state between two qubits $i$ and $%
k$ using the spin of the mobile electron as a bus qubit. The bias voltages
localize the electron at the appropriate positions, the ESR pulses (shown as
wavy arrows) induce SWAP operations between the stationary and the bus
qubit. }
\label{f.Remote}
\end{figure}

We consider an array of identical $^{31}P@C_{60}$ placed in a SWCNT as shown
in Fig. \ref{f.Remote}. As described in section \ref{s.readout}, it is
possible to exchange a quantum state between the stationary qubit and the
mobile electron spin. We may therefore use the mobile electron spin as a bus
qubit: If we want to transfer quantum information between the stationary
qubits $i$ and $k$, we first move the electron to qubit $i$, perform a SWAP
operation, using the appropriate ESR pulse sequence. We then move the mobile
electron to position $k$, perform a second SWAP operation, and back to
position $i$ for a third SWAP operation. This completes the SWAP$_{ik}$
operation.

The caged spin is well protected from external perturbations, but
decoherence experienced by the mobile electron spin may be faster.
Nevertheless, decoherence times of the order of 10 $\mu$s have been reported
\cite{Galland}, which would make it comparable to the relaxation times of
the N@C$_{60}$ in the peapods \cite{Toth} and be long enough to perform SWAP
operations as well as transport of the mobile electron fast compared to the
decoherence time.

\section{Feasibility}

The preceding sections presented a design for a scalable quantum computer.
For this concept to become reality, a number of engineering challenges have
to be overcome. Here, we discuss the design parameters that have to be
reached to make such a device useful.

\subsection{Addressing qubits}

Addressing the individual electronic spins requires that they are
distinguishable in frequency space. This is achieved by applying a magnetic
field gradient, which shifts the individual electron spin resonance. One
difficulty with the frequency-space addressing is that we must avoid
generating overlap between the hyperfine-split resonances of the individual
qubits. Furthermore, we have assumed that we work in the weak-coupling
limit, where the frequency differences between adjacent qubits are large
compared to the dipole-dipole coupling constant. While this is not a
necessary requirement \cite{Mah06}, it simplifies the design of quantum gate
operations.

\begin{figure}[h]
\centering\includegraphics[width=9cm]{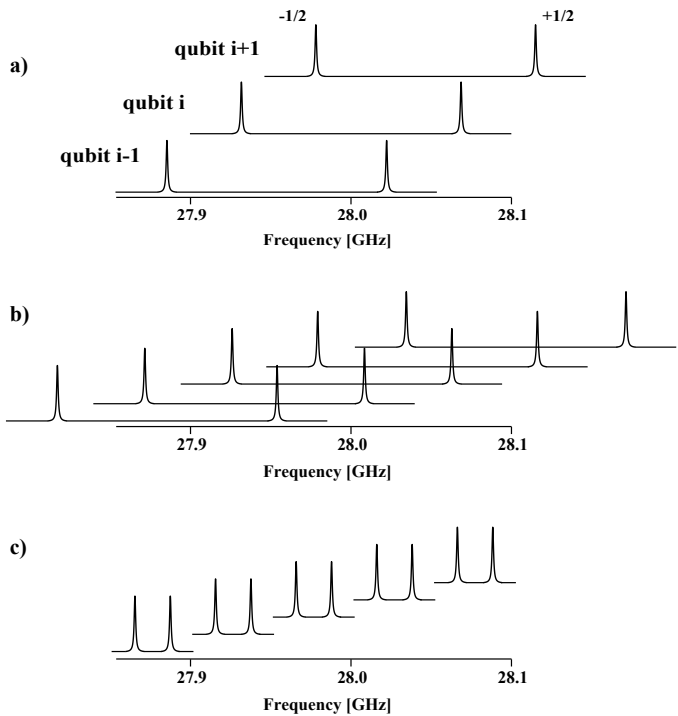}
\caption{Schemes for addressing electron spin qubits in a magnetic field
gradient. a) Three $^{31}$P@C$_{60}$ molecules, separated by $\Delta \Omega
=2\protect\pi \times 45$ MHz. b) Five $^{31}$P@C$_{60}$ molecules, separated
by $\Delta \Omega =2\protect\pi \times 55$ MHz. b) Five $^{15}$N@C$_{60}$
molecules, separated by $\Delta \Omega =2\protect\pi \times 55$ MHz.}
\label{f.sys}
\end{figure}

If we choose the separation between adjacent qubit molecules as $\Delta
z=2.9 $ $nm$, we obtain a dipole-dipole coupling constant of the order of 3 $%
MHz$. If we wish to stay in the week coupling limit, the separation between
the Larmor frequencies of neighboring electron spins should then be at least
30 $MHz$. As shown in Figure \ref{f.sys} a), this allows no more than 3-4
molecules in a row before the resonance frequencies of the higher frequency
lines of the hyperfine doublet start to overlap with the frequency range of
the lower-frequency lines of some other molecules. As shown in Figure \ref%
{f.sys} b), it is nevertheless possible to add more molecules without
generating overlap, provided the frequency separation between the qubits is
chosen correctly. An alternative approach is depicted in Figure \ref{f.sys}
c): Here, we assume that the difference between the Larmor frequencies of
adjacent molecules is large compared to the hyperfine splitting, thus
avoiding any spectral overlap between neighboring molecules. This choice of
parameters may be most suitable for $^{15}$N@C$_{60}$ molecules, where the
hyperfine splitting is significantly smaller than for $^{31}$P@C$_{60}$.

Compared to the electron spin resonance frequencies, the nuclear Larmor
frequencies are smaller by 3-4 orders of magnitude. In the field gradient
considered here, this would correspond to frequency differences of the order
of only 5 $kHz$ for $^{15}$N and 20 $kHz$ for $^{31}$P for the parameters
used above. Selective addressing of the nuclear spins thus requires that the
duration of the selective RF pulses should be at least several 100 $\mu $s,
which may be undesirable for fast gate operations. A good solution could be
the use of hard (short) RF pulses that excite all nuclear spins. As
described in the previous sections, each SWAP operation between the electron
and nuclear spins includes an even number of $\pi $ pulses acting on the
nuclear spins. For the nuclear spins of the passive qubits, the SWAP
operations result in a NOP, as required.

\subsection{Experimental challenges}

While the quantum architecture discussed here appears attractive and
physically feasible, its implementation faces several formidable challenges.
Several proposals have been put forward for injecting and moving polarized
electrons on SWCNT \cite{kou,Gun,ben}. Turning those into useful devices for
quantum computing will require precise control by electrodes that allow one
to move the mobile electrons between the required positions without
affecting the quantum mechanical superposition states of their spin degrees
of freedom. Currently, there are no date available about the relevant
decoherence times for the spins of these mobile electrons. The closest
references are from electrons in SWCNT-based single-electron devices \cite%
{Pea2}. Since the relevant decoherence mechanism is not yet fully
understood, it is difficult to extrapolate. Depending on the mechanism and
parameters, it may be possible to use dynamic decoupling techniques \cite%
{viola, Bang} to reduce the decoherence.

In the main part of the paper, we have considered exchange of information
between different parts of the quantum register by direct dipolar
interactions or with the help of a mobile electron spin acting as a bus
qubit. If these approaches prove too difficult to scale to many-qubit
quantum registers, it might be worth considering a distributed quantum
computer based on many small peapod quantum registers. They could be
connected via the Bell-state analyzer discussed in Ref. \cite{Bee}, which
are designed to entangle two electron spins. As shown in \cite{Yang},
passing the mobile electrons from different C$_{60}$@SWCNTs through the
Bell-state analyzers could entangle the caged electron spins in different C$%
_{60}$@SWCNTs, as required for building a quantum network \cite{kimble}. To
build the Bell-state analyzers, however, we have to develop beam splitters
for the electrons and suitable charge detectors. Although there has been
some progress in these aspects \cite{henli, elzer}, these devices are not
yet available \cite{elzer}.

Additional challenges are associated with the peapod itself: depending on
the system parameters, the distances between the molecules may vary, and the
molecules may be arranged in zig-zag form rather than in a straight line
\cite{ben}. In natural abundance, the 1\% fraction of carbon nuclei that
carry a nuclear spin ($^{13}$C) will contribute to the decoherence in the
system. Clearly, this effect can be reduced by using isotopically pure
material, as has been demonstrated, e.g., in diamond \cite{pure}. Clearly,
these challenges are difficult but may be overcome eventually. It will be
interesting to watch progress in this area.

\section{CONCLUSION AND ACKNOWLEDGMENTS}

In summary, we have proposed a scheme for performing universal quantum
computation in a C$_{60}$@SWCNT (peapod) system. We have discussed how to
efficiently implement the quantum logical gate operations required for
universal quantum computation. Transfer of information between qubits was
considered by direct dipole-dipole couplings or by using a mobile electron
spin as a bus qubit. While our scheme cannot be realized with today's
technology, it appears possible that the currently existing obstacles can be
overcome as nanotechnology makes further progress.

This work is supported by NNSF of China under No. 10774163 and by the Robert
Bosch Stiftung.

\end{document}